\documentclass[aps,twocolumn,showpacs,floats,superscriptaddress]{revtex4}
\usepackage{graphicx}
\usepackage{amssymb,amsmath}
\begin{document}

\title{Superfluid-Insulator and Roughening Transitions in Domain Walls}

\author{\c{S}.G. S\"{o}yler}
\affiliation{Department of Physics, University of Massachusetts,
Amherst, MA 01003, USA}

\author{B. Capogrosso-Sansone}
\affiliation{Department of Physics, University of Massachusetts,
Amherst, MA 01003, USA}

\author{N.V. Prokof'ev}
\affiliation{Department of Physics, University of Massachusetts,
Amherst, MA 01003, USA}  \affiliation{Russian Research Center ``Kurchatov
Institute'', 123182 Moscow, Russia}

\author{B.V. Svistunov}
\affiliation{Department of Physics, University of Massachusetts,
Amherst, MA 01003, USA}
\affiliation{Russian Research Center ``Kurchatov Institute'',
123182 Moscow, Russia}

\begin{abstract}
We have performed quantum Monte Carlo simulations to investigate
the superfluid behavior of one- and two-dimensional interfaces
separating checkerboard solid domains. The system is described by
the hard-core Bose-Hubbard Hamiltonian with nearest-neighbor
interaction. In accordance with Ref.~\cite{Burovski}, we find that
(i) the interface remains superfluid in a wide range of
interaction strength before it undergoes a superfluid-insulator
transition;  (ii) in one dimension, the transition is of the
Kosterlitz-Thouless type and is accompanied by the roughening
transition, driven by proliferation of charge-$1/2$
quasiparticles; (iii) in two dimensions, the transition belongs to
the 3D $U\left(1\right) $ universality class and the interface
remains smooth. Similar phenomena are expected for domain walls in
quantum antiferromagnets.

\end{abstract}

\maketitle

\parindent  8.mm

\section {Introduction}

Strongly correlated quantum lattice systems represent one of the
most exciting and active fields in condensed matter physics. On
the fundamental physics side, they provide a rich playground to
investigate quantum phase transitions and study new exotic states
of matter (for a review see e.g. \cite{Bloch} and as an example of
more exotic systems see e.g. \cite{exotic_1, exotic_2}). On the
experimental side, the possibility of trapping bosons in optical
lattices in a highly controllable manner makes such systems good
candidates for applications in a variety of different fields such
as quantum communication, computing, and precision measurements
\cite{Jaksch, Duningham, Rodriguez}. In particular, recently there
has been a great interest toward studying trapped cold polar
molecules for which the possibility of controlling long-range
dipole-dipole interactions opens up the way to realization of
novel quantum phases and possible use of this system for above
mentioned applications (see \cite{polar_molecules1} and references
therein).

Studies of quantum phases and transitions between them are mostly
confined to well defined geometries and lattice types in a given
dimension. It is, however, important to recognize that
low-dimensional quantum systems can also emerge in the form of
extended defects in a higher dimensional regular structure. Domain
walls, grain boundaries and dislocations are the most prominent
examples of such systems. Recently, stimulated by the observation
of non classical moment of inertia in solid He-4 \cite{NCMI},
superfluid properties of defects in crystals were looked at in
Refs.~\cite{Burovski,He-gb,Screw disl.}. In particular, it has
been shown that it is possible to have lower dimensional
superfluid phases emerging on topologically frustrated defects in
solid $^4$He.
\\ \indent For the scope of this work we are particularly interested
in the results of Ref.~\cite{Burovski}, where the authors studied
universal properties of the superfluid-insulator (SF-I) transition
in interfaces separating two checkerboard (CB) domains by
simulating the classical (d+1)-dimensional bond-current model
\cite{bond-current}. Here we briefly summarize their results. It
has been shown that the grain boundary remains superfluid in a
large range of parameters before undergoing the SF-I transition.
In one dimension, the transition is of the Kosterlitz-Thouless
(KT) type. An argument explaining why one should expect a
roughening transition to happen simultaneously to the SF-I
transition has been given: both transitions are driven by
proliferation of one and the same quasiparticle carrying charge
$1/2$. In two dimensions, however, the interface remains smooth
and the transition belongs to the 3D $U\left(1\right) $
universality class.
\\ \indent At the qualitative level, the model studied in Ref.~\cite{Burovski}
works for all systems of the same universality class. However, it
can not be used for making quantitative predictions regarding
realistic quantum bosonic Hamiltonians. These predictions are one
of the main goals of our paper. In addition, we present a direct
evidence for the fact that topological excitations carry
charge-$1/2$, and \emph{quantitatively} discuss the connection
between roughening and SF-I transitions.

We are interested in studying grain boundaries in the CB solid
which may be created in the system of hard-core cold bosons taken
across the SF-CB transition. A possible experimental realization
is represented by cold polar molecules in an optical lattice. In
the interesting experimental regime, interparticle distances are
such that double occupancy is strongly suppressed (giving rise to
hard core bosons), while the nearest-neighbor interaction can be
tuned via external electric fields
\cite{polar_molecules1,polar_molecules2}.
\\ \indent The hard-core extended Bose-Hubbard Hamiltonian reads:
\begin{equation}
H= -t\sum_{<ij>} \left( a^{\dag}_i\,a_j + h.c.\right) +
V\sum_{<ij>}n_in_j -\sum_i \mu_i n_i\; , \label{hamiltonian}
\end{equation}
where $a^{\dag}_i (a_i)$ is the bosonic creation (annihilation)
operator, $t$ is the hopping matrix element, $V$ is the nearest
neighbor repulsion, and $\mu$ is the chemical potential. On a simple
cubic/square lattice the model (\ref{hamiltonian}) is equivalent to
the spin-1/2 XXZ antiferromagnet with exchange couplings
$|J_x|=|J_y|=2t$, $J_z=V$ and magnetic field $h=zV/2-\mu$, where $z$
is the number of nearest neighbors. This correspondence makes our
work equally relevant for studies of domain walls in spin systems,
especially in the limit of strong anisotropy (or small domain wall
width when the continuous approximation breaks down). The zero
magnetic field case, i.e. $\mu=zV/2$, corresponds, in bosonic
language, to half-integer filling factor. At half filling, the
ground state of model (\ref{hamiltonian}) features the SF and CB phases only. The SF phase,
characterized by broken \emph{U}(1) symmetry, corresponds to the
easy-plane antiferromagnet with the order parameter $\Psi=S_x+iS_y$,
while the CB order, characterized by broken $Z_{2}$ symmetry,
corresponds to the easy-axis antiferromagnet with long-range
correlations in $S_z$ (here $S_{x,y,z}$ describe components of the
N\'{e}el vector). While at a generic filling factor the system
undergoes phase separation into SF and CB phases
\cite{ground_state_phase_1, ground_state_phase_2}, at half filling
the SF-CB transition happens at a special higher-symmetry point,
$V=2t$, where the Hamiltonian (\ref{hamiltonian}) features an
\emph{O}(3) symmetry. In the spin model it corresponds to the
Heisenberg point. At the Heisenberg point the N\'{e}el vector lives
on a sphere and can point in any direction; it simply rotates from
equator to the pole when going from the SF to the CB phase giving
rise to the discontinuous (in terms of order parameters) SF-CB
transition.

In what follows we consider a system initially prepared with two
large solid domains with $S_z=M$ and $S_z=-M$ respectively,
separated by a domain wall. To achieve this we consider a system of
size $L_x=N+1, L_y(=L_z)=N$ where N is an even integer, with
periodic boundary conditions (PBC). For the easy-axis Hamiltonian,
this results in a frustrated CB where two atoms can avoid occupying
nearest neighbor sites everywhere but in the domain wall layer. Our
goal is to study properties of the resulting $(d-1)$-dimensional
interface embedded in a $d$-dimensional solid using quantum Monte
Carlo simulations. We focus on the case of zero magnetic field (half
filling factor), i.e. the chemical potential is fixed to $\mu=zV/2$,
and show that the interface remains SF (has gapless magnons, in
magnetic language) well beyond the bulk SF-CB transition, for a wide
range of $V/t$.

Another interesting question concerns the roughness of the
interface (for a brief review of roughening transition see e.g.
\cite{roughening}). To be specific, an interface (viewed on large
scale as a membrane) is said to be smooth if its mean square
relative displacement, see Eq.~(\ref{delta_x}) below, is finite in
the thermodynamic limit. On the contrary, if the latter diverges
with the system size, an interface is said to be rough. In this
work we present numerical results, taken across the SF-I
transition, of the mean square relative displacement for several
system sizes.

The paper is organized as follows. In Subsection II A we present
results for the 1D interface and show that the superfluid and
roughening transitions happen simultaneously. In Subsection II B we
consider a 2D interface and calculate the critical point for the
SF-I transition. We briefly summarize our results in the Conclusion.

\section {Results}
Our simulations are based on the Worm algorithm path integral
approach \cite{worm algorithm}, which allows efficient sampling of
the many-body path winding numbers $\left\langle W_{\alpha}^2
\right\rangle$ in imaginary time and space directions. In the path
integral representation winding numbers have a simple geometrical
meaning: $W_{\tau}=N$ is the number of particles in the system,
and $W_{x,y,z}$ describe how many times the many-body trajectory
loops around the system with periodic boundary conditions in
directions $\hat{x}$, $\hat{y}$ or $\hat{z}$. Thus, winding number
fluctuations in temporal and spatial directions define the
compressibility and superfluid stiffness, respectively. Since we
consider the easy-axis Hamiltonian at $V>2t$ and half integer
filling, i.e. the system is in the insulating CB phase, the
superfluidity is restricted to the plane of the grain boundary. As
a consequence, we expect a nonzero value for spatial winding
numbers only along directions parallel to the grain boundary. In
the following we show that the superfluid response persists up to
a critical value $(V/t)_c$, at which the interface undergoes the
SF-I transition. In 1D interfaces this transition is of the
Kosterlitz-Thouless type \cite{KT} and is accompanied by
roughening. In 2D the transition belongs to the 3D U(1)
universality class and the interface remains smooth.
\\ \indent In order to calculate critical points (Fig.~\ref{G} and
Fig.~\ref{critical_point_3D} below) we have performed simulations in
a square system, i.e. $L_{\tau}\approx L_{y,z}=L$, where
$L_{\tau}=c\beta$. Here $c=\sqrt{\rho_{s}/\kappa}$ is the sound
velocity. In practice, for each system size we choose $\beta$ such
that $\left\langle W_{\tau}^2 \right\rangle\sim\left\langle
W_{y,z}^2 \right\rangle$. This requirement implies $c\approx
L/\beta$, with the inverse temperature $\beta$ scaling with the
system size. The last expression for $c$ comes from the dependence
on winding numbers of the superfluid stiffness
$\rho_{s}=\frac{L^{2-d}\langle\overrightarrow{W}^2\rangle}{\beta
d}$, where $d$ is the system dimensionality and
$\langle\overrightarrow{W}^2\rangle=\sum_{i=1,d}\;\langle{W_i^2}\rangle$,
and compressibility
$\kappa=\frac{\beta\langle{W_{\tau}^2}\rangle}{V}$, where $V$ is the
volume.

\subsection {One dimensional interface}
Here we consider a system with $L_x=N+1$, $L_y=N$, which implies a
grain boundary along the $\hat{y}$-direction. We start our study at
$V/t=2.3$, slightly above the Heisenberg point $V/t=2$. We recall
that winding number fluctuations in the superfluid phase are
described by the gaussian distribution (see e.g. \cite{sf_density}),
which in $d=1$ takes the form $P(W_y^2) \propto \exp( -LW_y^2/2\beta
\rho_s)$. Similarly for winding numbers in imaginary time direction
(particle number fluctuations) $P(W_\tau^2) \propto \exp(-\beta
W_\tau^2/2L\kappa)$. In $d=1$ the distributions are essentially
discrete and the best way to extract superfluid stiffness and
compressibility from the simulated distributions is by considering
the following ratios:
\begin{equation}
 \rho_s^{-1}= 2\frac{\beta}{L} \ln \left[ \frac{2
 P(W_y=0)}{P(W_y=-1)+P(W_y=+1)} \right] ,\label{rho_s}
\end{equation}
and
\begin{equation}
\kappa^{-1}= 2\frac{L}{\beta} \ln \left[ \frac{2
P(W_\tau=0)}{P(W_\tau=-1)+P(W_\tau=+1)} \right]\; .\label{kappa}
\end{equation}

\begin{figure}[t]
\centerline{\includegraphics[angle=0,scale=0.55] {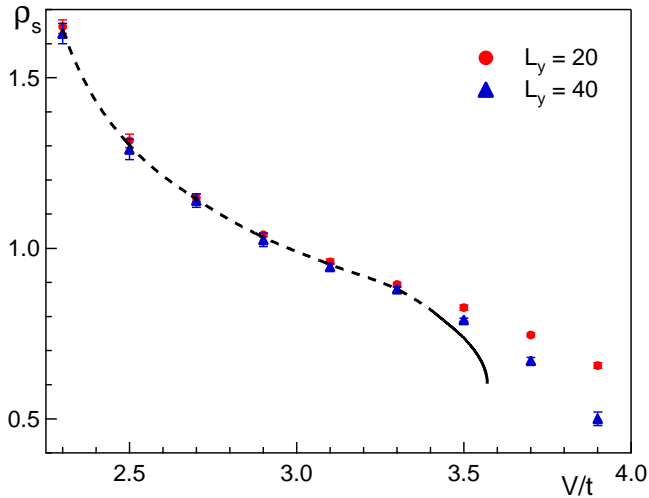}}
\caption{(Color online) Superfluid stiffness  ($\rho_s$) vs
interaction strength (V/t) for system sizes $L_y=20$ (circles) and
$L_y=40$ (triangles) at zero temperature. The dashed line is to
guide an eye, and the solid line represents $\rho_s$ in the
thermodynamic limit, from finite size scaling. Error bars are
within the symbol size.} \label{SF_1D}
\end{figure}

We find that $W_x^2=0$ in our system sizes, ensuring that non-zero
values for $W_y^2$ are due to grain boundary only. In
Fig.~\ref{SF_1D} we plot the superfluid stiffness as a function of
the interaction strength $V/t$, for system sizes $L_y=20$
(circles) and $L_y=40$ (triangles). The temperature has been
chosen much smaller than the finite size energy gap, so that the
system is effectively at zero temperature. The grain boundary
remains superfluid, and, within statistical error bars, we do not
see any size effects up to $V/t\gtrsim 3.5$. Finite-size effects
indicate that we are approaching the SF-I transition in the
interface.

\begin{figure}[t]
\centerline{\includegraphics[angle=0,scale=0.48] {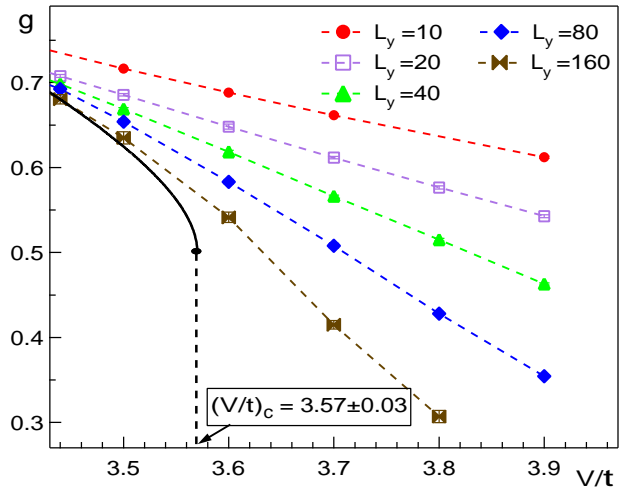}}
\caption{(Color online) Luttinger liquid parameter \emph{g} as a
function of $V/t$. (Error bars are within the symbol size). Dashed
lines are to guide an eye; the solid line is the result of
extrapolation to the infinite system size based on the
Kosterlitz-Thouless theory. Simulation results are consistent with
the universal jump at $g_c=0.5$. The critical value is at
$(V/t)_c=3.57\pm0.03$.} \label{G}
\end{figure}

The 1D interface forms a Luttinger liquid and the quantum phase
transition to the insulating state is of the Kosterlitz-Thouless
type, characterized by the universal jump at $g=g_c=2/m^2$, where
$g=\pi\sqrt{\rho_s\kappa}$ is the dimensionless Luttinger liquid
parameter (see e.g. \cite{g-luttinger}) and $1/m$ is the fractional
filling factor. In Fig.~\ref{G}, we show $g$ as a function of $V/t$
for various system sizes (data are taken in square systems).
Clearly, simulation results are fully consistent with the universal
jump at $g_c=0.5$, which implies an effective filling factor $1/2$
in the grain boundary. The only logical explanation for fractional
filling factor when translation symmetry in the bulk is broken with
doubling of the unit cell volume, is to assume that the SF boundary
is rough. Rough interface effectively averages the bulk potential
and thus retains the original translation symmetry of the lattice.
At the microscopic level, superfluidity and roughening are both
linked to the proliferation of spinon excitations, and we provide
additional evidence for this explanation below. In the figure, the
solid line is the result of finite-size scaling based on the
Kosterlitz-Thouless renormalization group (RG) flow \cite{KT}. The
integral form of the equations reads:
\begin{equation}
\int_{g(L_1)/g_c}^{g(L_2)/g_c}\dfrac{dt}{t^2(\ln(t)-\xi)+t
}=4\ln(L_1/L_2),\label{g_equation}
\end{equation}
where $\xi$ is a system size independent parameter. Using numerical
data for $g(L)$ the integral can be solved numerically and
$\xi(V/t)$ can be evaluated for each pair of sizes ($L_1$,$L_2$). In
the critical region one expects $\xi$ to be a linear function of the
interaction potential. At the critical point, Eq.(\ref{g_equation})
is satisfied by $\xi=1$.

\begin{figure}[t]
\centerline{\includegraphics[angle=0,scale=0.6] {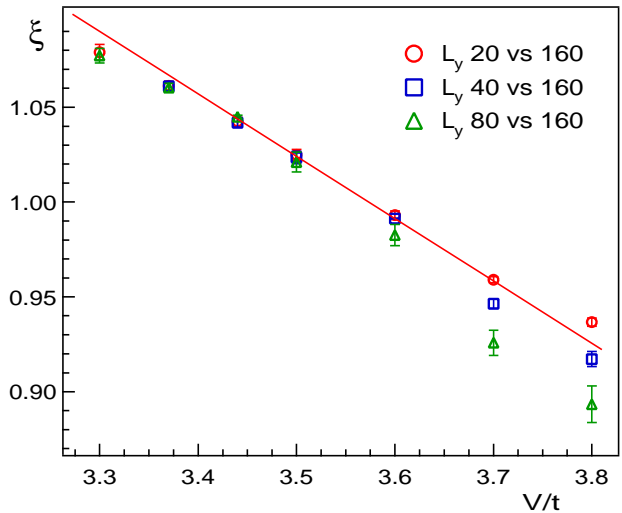}}
\caption{(Color online) Parameter $\xi$ for different pairs of
system sizes  $(L_1,L_2)$ as a function of (V/t). The solid line
is a linear fit. The transition point is determined from the $\xi
(V/t)=1$ condition.} \label{ksi_fig}
\end{figure}

In Fig.~\ref{ksi_fig} we show the solution of Eq.~(\ref{g_equation})
for three different pairs of ($L_1$,$L_2$). The data show a good
collapse in the vicinity of the critical point
$(V/t)_c=(3.57\pm0.03)$. The $L\rightarrow\infty$ limit of
Eq.~(\ref{g_equation}) yields
$(g(V/t)/g_c)(\ln(g(V/t)/g_c)-\xi(V/t))=-1$; its solution determines
the thermodynamic value of the Luttinger liquid parameter $g$
indicated by the solid line in Figs. \ref{G} and \ref{SF_1D}.

\begin{figure}[t]
\centerline{\includegraphics[angle=0,scale=0.42]
{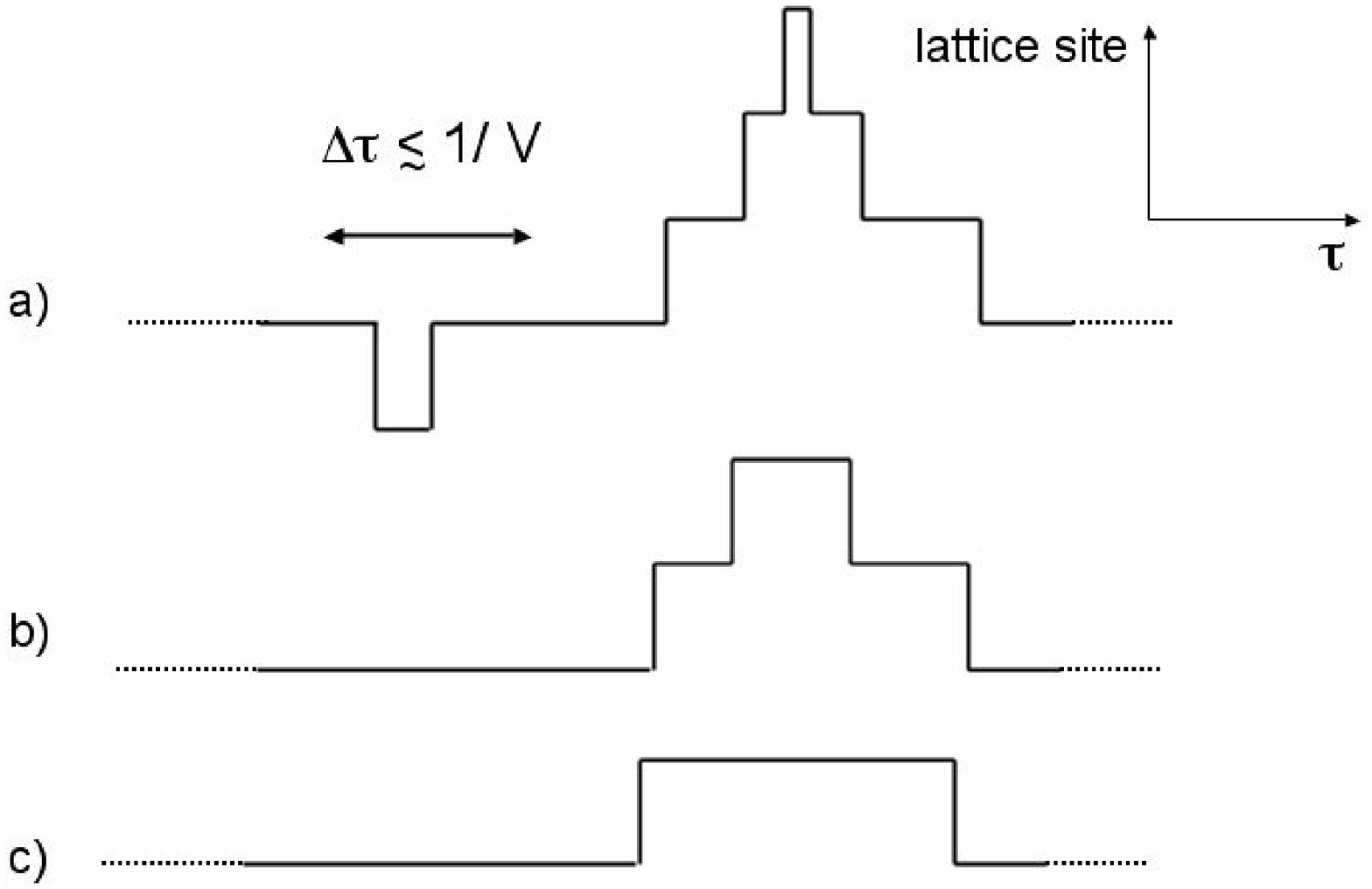}} \caption{Here we show an example on how we
``filter" the original trajectory a) by erasing hopping events,
happening on a time scale $\Delta\tau\lesssim 1/V$, which connect
the same nearest neighbor sites. Trajectories b) and c) show the
result of `first and second filtering stage' respectively. We stop
the process at the second stage.} \label{filtering}
\end{figure}

We can explicitly verify that topological excitations driving the
transition in 1D carry an effective charge of $1/2$. As it has
been argued by Burovski \emph{et al.}, the propagation of kinks,
or spinons, is achieved by shifting particles along the grain
boundary. A single particle hopping event moves the grain boundary
in the transverse direction and shifts the kink by two lattice
steps. It means that in the presence of a gauge field a kink going
around the system will accumulate half the gauge phase an ordinary
particle will, i.e. its effective quasiparticle charge is 1/2.\\
\indent In order to show that this is the case we  measure how
winding number fluctuations develop in imaginary time. The
measurement is done for the insulating domain wall when winding
number fluctuations are rare and one can study individual events.
More specifically, we monitor $W_y^2$ and pick configurations with
$W_y^2=1$. To suppress noise originating from quantum fluctuations
in the solid bulk, we ``filter" the trajectory by erasing hopping
events which connect the same nearest neighbor sites and follow
each other in imaginary time (see Fig.~\ref{filtering}).\\ \indent
\begin{figure}[t]
\centerline{\includegraphics[angle=0,scale=0.52]
{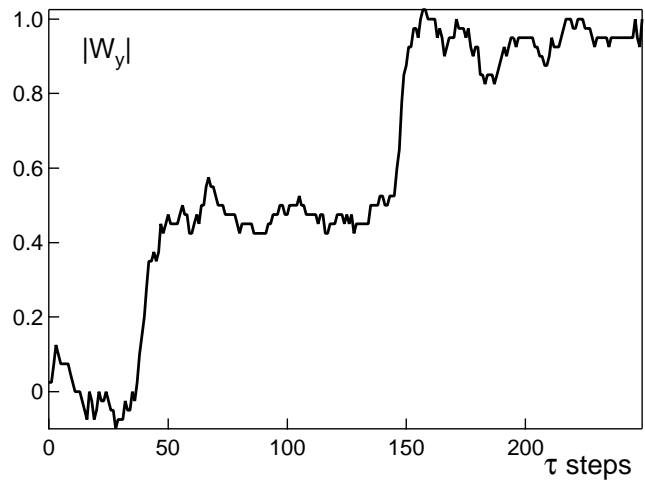}} \caption{The development of the non-zero
winding number along the grain boundary in imaginary time. The
data refer to a system of size $L_y=20$ and $V/t=4.2$. The
presence of steps implies that $W$ fluctuations happen in the form
of instantons, i.e. the transition time from one ground state
(degenerate in the thermodynamic limit) to another is a rare and
fast virtual event. The step amplitude $\approx 1/2$ proves the
fact that relevant elementary excitations carry charge $1/2$. }
\label{instanton}
\end{figure}
Figure \ref{instanton} shows an example of the winding number
trajectory for a system with $L=20$ and $V/t=4.2$. The first step
describes system transition to an equivalent ground state obtained
by shifting interface particles once, i.e. with the domain wall
shifted by one unit length in the transverse direction. At this
point the system has two choices: going back to the initial state
or making another transition in the same direction and completing
the full winding number (which has to be integer). This latter
case is shown in Fig.~\ref{instanton} because we consider a
configuration with $W_y^2=1$. As one moves closer to the
transition point spinon excitations become more frequent and start
overlapping in imaginary time making the whole picture less
transparent for analysis.

Next we would like to discuss the connection between SF and
roughening more quantitatively. In order to do so we have calculated
the mean square displacement $\langle\Delta x^2\rangle$ of the grain
boundary profile, where the average is done in both, imaginary time
and space:
\begin{equation}
\left\langle \Delta x^2 \right\rangle = \left\langle
\int_{0}^{L_y}\int_{0}^{\beta} \Delta x^2(y,\tau) d\tau
dy\right\rangle \;.\label{delta_x}
\end{equation}
Here $\Delta x(y,\tau)$ is the instantaneous position of the center
of the wall (recall that the interface is along the $y$ direction),
calculated from the center of the grain boundary at $(\beta
L_y)^{-1} \int \int x(y,\tau) d\tau dy $. To determine $x(y,\tau)$,
we calculate the difference between the ``instantaneous" densities
of two consecutive sites along the $x$ direction, $\delta
n(i,\tau)=|n(i,\tau)-n(i+\hat{x},\tau)|$. Note that by
``instantaneous" density, $n(i,\tau)$,
 we mean an average over some time interval $\Delta\tau$ such that
$1/V\ll\Delta\tau\ll\beta$, in order to eliminate zero point
fluctuations. Within the checkerboard solid, $\delta
n(i,\tau)\approx1$. In the grain boundary, instead, one can have
consecutive sites either both occupied or empty (with a fractional
value of $n(i,\tau)$ in the superfluid state of the boundary).
Practically, for any given $y$, we start from the first lattice
site and determine $x(y,\tau)$ from $\min(\delta n(i,\tau))$.\\
\indent
\begin{figure}[t]
\centerline{\includegraphics[angle=0,scale=0.50] {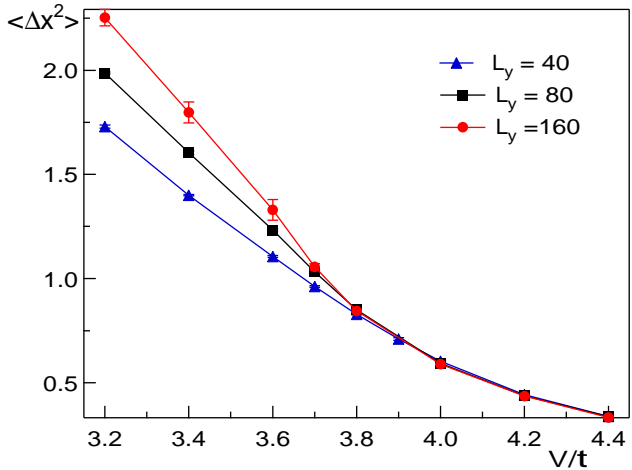}}
\caption{(Color online) The mean square displacement (measured in
units of lattice spacing) of the grain boundary profile as a
function of (V/t). In the insulating phase the data for different
system sizes collapse, as expected for a smooth interface. Data
start splitting in the vicinity of the critical point. In the
superfluid state the interface is rough and $\langle\Delta
x^2\rangle$ depends on the system size. } \label{fdelta_x}
\end{figure}
By definition, $\langle\Delta x^2\rangle$ is expected to be $\sim 1$
and system size independent for a smooth interface. For a rough
interface, however, $\langle\Delta x^2\rangle$ diverges with the
system size. Our results are consistent with this expectation as
Fig.~\ref{fdelta_x} shows. Beyond the transition to the insulating
state, results for different system sizes overlap within statistical
errors. On the superfluid side, instead, $\langle\Delta x^2\rangle$
increases with the system size $L$.

\subsection {Two dimensional interface}
\begin{figure}[t]
\centerline{\includegraphics[angle=0,scale=0.47]
{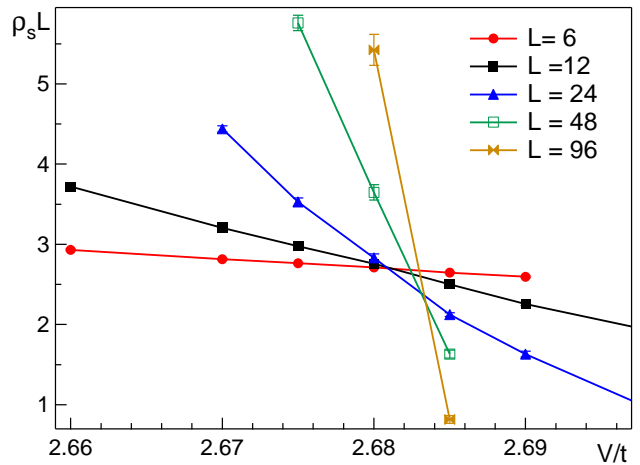}} \caption{(Color online) Finite size
scaling for the superfluid stiffness of a 2D interface. From the
intersection of curves we estimate $(V/t)_c=2.683\pm0.003$. Solid
lines are to guide an eye. } \label{critical_point_3D}
\end{figure}
Here we present results referring to a 3D system, i.e. 2D interface,
at half integer filling factor. In this case the phase transition
belongs to the $U(1)$ universality class in three dimensions,
indicating a SF---Mott insulator transition at integer filling
factor. As discussed in Ref.~\cite{Burovski}, the bulk acts as an
effective periodic potential which doubles the primitive cell in the
interface (if it remains smooth at the transition), bringing the
filling factor from $1/2$ to $1$. In Fig.~\ref{critical_point_3D} we
show the finite size scaling of the superfluid stiffness obtained
from $\rho_s=\left\langle W^2_y + W^2_z\right\rangle / \beta$
(recall that the transverse direction is $x$), with $\beta$ scaling
with the system size. From the intersection of scaled curves
referring to different system sizes we obtain the critical point at
$(V/t)_c=2.683\pm0.003$.

In 3D, we did not observe quasiparticles carrying fractional
charge $1/2$. Hopping of single particles along the interface
remains a local fluctuation; to shift the wall one has to create a
macroscopic line defect (``atomic step") which is energetically
expensive. Since the energy barrier between equivalent ground
states increases with system size the interface remains smooth at
$T=0$.

\section{Conclusions}

To summarize, we studied superfluid-insulator transitions in 2D and
1D domain walls in the bosonic checker-board solid. In both cases
domain walls remain superfluid well past the bulk SF-CB transition.
In 1D the SF-I transition in the wall is intrinsically linked to the
interface roughness since both properties are due to the
proliferation of fractionally charged spinon excitations.

The minimal description of the system is given by the hard-core
extended Bose-Hubbard Hamiltonian. Because on the bi-partite lattice
the latter can be exactly mapped onto a spin 1/2 XXZ antiferromagnet
model, the results presented here are also relevant to spin systems
(for other lattice types the equivalent spin model is ferromagnetic
in the {\it xy}-plane and antiferromagnetic along the {\it z}-axis).
One can imagine creating domain walls in the system of ultra cold
bosons in the process of fast ramping of the optical potential with
several solid seeds nucleated at different locations in the trap.
Clearly, the dynamics of grain boundaries and sample ``annealing"
will crucially depend on the the SF/roughening transitions. It might
be also possible to observe signatures of lower-dimensional
coherence in absorption images and interference experiments where
none are expected for the ideal insulating bulk state.

The work was supported by the National Science Foundation under
Grants PHY-0426881 and  PHY-0653183.


 \end{document}